The manuscript is submitted to *Acta Materialia*

**Title**

# Role of stress-assisted martensite in the design of strong ultrafine-grained duplex steels

**Author**

Hung-Wei Yen [a, b, *], Steve Woei Ooi [c, *], Mehdi Eizadjou [a, b], Andrew Breen [a, b], Ching-Yuan Huang [d], H. K. D. H. Bhadeshia [c], and Simon P. Ringer [a, b, *]

**Affiliation Address**

a. School of Aerospace, Mechanical, Mechatronic Engineering, The University of Sydney, NSW 2006, Australia
b. Australian Centre for Microscopy and Microanalysis, The University of Sydney, NSW 2006, Australia
c. Department of Materials Science and Metallurgy, University of Cambridge, 27, Charles Babbage Road, Cambridge CB3 0FS, U.K.
d. Iron & Steel Research & Development Department, China Steel Corporation, Kaohsiung, 81233, Taiwan, R.O.C.

***Corresponding Author:**

Hung-Wei Yen e-mail: homer.yen@sydney.edu.au

Steve Woei Ooi e-mail: swo23@cam.ac.uk

Simon P. Ringer e-mail: simon.ringer@sydney.edu.au

**Abstract**

This work explains the occurrence of transformation-induced plasticity via stress-assisted martensite, when designing ultrafine-grained duplex steels. It is found that, when the austenite is reduced to a fine scale of about 300 nm, the initial deformation-induced microstructure can be dominated by parallel lamellae of ε martensite or mechanical twinning, which cannot efficiently provide nucleation sites for strain-induced martensite. Hence, α′ martensite nucleation occurs independently by a stress-assisted process that enhances transformation-induced plasticity in ultrafine-grained austenite. This metallurgical principle was validated experimentally by using a combination of transmission Kikuchi diffraction mapping, transmission electron microscopy, and atom probe microscopy, and demonstrated theoretically by the thermodynamics model of stress-assisted martensite.

## 1 Introduction

Strong steels that possess the combination of properties necessary for the manufacture of automobiles are the subject of intense research and development because of the increasing need for fuel efficiency, emissions reduction and safe design [1]. It is recognised, nevertheless, that there are driving forces for further alloy development in this space, and a particular focus has been on the development of variants that are strong and ductile [2-4]. Extreme grain-refinement in general leads to strengthening [5, 6], but the ductility is dramatically reduced [7, 8]. When the boundaries are so closely spaced, the dislocations sink into the boundaries or mutually annihilate causing the uniform ductility to become vanishingly small because of the lack of a prominent work hardening mechanism. In searching for strong but ductile ultrafine-grained steels, source of mobile dislocation [9] or other mediate of plasticity [2, 4, 10, 11] will be required .

There are three essential plastic deformation mechanisms in austenite at ambient temperature [12]: dislocation slip, the shape deformation accompanying the ε or α′ martensitic transformation, and mechanical twinning. Tailoring the chemical and structural characteristics of the austenite can invoke the operation of particular modes of plastic deformation, as required. This is important when designing fine grained steels [2, 4, 10, 11] and those that are nanocrystalline [13, 14], in order to avoid the well-known loss of ductility through necking shortly after yield. In these steels, other plastic modes such as martensitic transformation or mechanical twinning are essential in order to enable strain hardening. This idea has been utilised in the recent development of fine-grained duplex (austenite-ferrite or austenite-martensite) steels with 5-7 wt. % of manganese [2, 4, 15]. Those duplex microstructures benefit from the retained austenite, which can transform into martensite during deformation, thereby preventing plastic instabilities. However, such alloys [2, 4, 10, 15] generally yield at stresses well below 1 GPa, but ultrahigh yield strength will be also required especially for the various intrusion applications or hole-expanded components in the automotive industry [1]. Moreover, some of the medium manganese steels require prolonged heat treatments so as to ensure solute partitioning in order to stabilise the austenite [16, 17], making manufacturing process time and energy consuming.

The present work was undertaken to resolve these difficulties by designing a new alloy containing a manganese concentration of 11 wt. %. This special alloy can be very strong and ductile due to its ultrafine-grained duplex structure, produced by a short annealing. Particular attention was devoted to elucidating the metallurgical principles of TRIP effect when the grain size is extremely small. The tensile mechanical behaviour is presented together with the microstructural details which were obtained using scanning electron microscopy (SEM) based transmission Kikuchi diffraction (TKD), transmission electron microscopy (TEM), and the atom probe microscopy (APM). These techniques provide direct evidence and data for the thermodynamics model, indicating an enhanced stress-assisted TRIP effect, which has not been reported in steels of this grade.

## 2 Alloy Design

The steel composition developed in this research is presented in Table 1. The steel was prepared by vacuum induction melting, and cast into ingots of size 80 × 30 × 12 $mm^3$. The alloy was designed to possess a very fine-grained austenite-ferrite duplex microstructure that could be achieved by a combination of cold rolling and intercritical annealing (i.e.) heat treatment in the austenite and ferrite two phase field. There are several aspects to the novelty in the proposed composition. Firstly, the steel possesses a much lower carbon content than the steels reviewed in the previous section, Table 1. This ensures that cold rolling down to sheet as thin as 1.2 mm gage thickness can be achieved from the martensitic state that predominates after hot rolling. The low carbon content also suppresses the formation of cementite, and so enables a larger temperature window for intercritical annealing. The 10-11 wt. % of manganese increases the stability of the austenite generated during intercritical annealing and yet is substantially lower than that used in the TWIP steels. The addition of vanadium provides for grain size control, and that of molybdenum for mitigating embrittlement by phosphorus and similar impurities, given the susceptibility that strong steels have to intergranular failure. Our aim was to achieve a duplex microstructure having equal volume fractions of austenite and ferrite so as to enhance the materials capacity for work hardening during deformation of the austenite.

The ingot was soaked at 1200 °C for 1 h and then hot-rolled to 6 mm thick plate followed by air cooling to room temperature. The plate was then reheated to 1200 °C for 30 min and then hot rolled to 2.8 mm thick strip, with a finish rolling temperature of 850 °C. This hot-rolled strip was then air cooled to room temperature. After removing the oxide layer on the steel surface via pickling, its thickness was reduced from 2.4 mm to 1.2 mm via a final cold rolling (CR) step, over eight passes at room temperature.

Figure 1a charts the equilibrium phase fractions by weight as a function of temperature, calculated using Thermo-Calc [18] with the TCFE7 database. The equilibrium weight fractions of austenite at the temperatures of 550 and 610 °C were 0.52 and 0.80, respectively. Considering the small difference in atomic density between austenite and ferrite, the volume fractions were also close to these values. Our target austenite volume fraction of 0.5 can therefore be achieved via an intercritical anneal at either temperature. At 610 °C, it would be necessary to interrupt the heat treatment prior to equilibrium, whereas at 550 °C, it would be necessary to arrest the heat treatment close to equilibrium. What is of crucial significance, however, is that these two intercritical annealing temperatures can grow austenite with significantly different characteristics. The corresponding austenite transformation kinetics are set out in Fig. 1b on the basis of dilatometry measurements made during annealing at 550 and 610 °C using the methods reported in [19, 20]. The kinetic analysis of the austenite transformation reveals that approximately 100 min is required to reach the required $V_\gamma$ =0.5 at 550 °C. In contrast, only 370 s is required at 610 °C. To enable the use of sufficiently large samples for tensile testing, strip samples were annealed in a box furnace. Given the relatively slow heating rate (approximately 3-4 °C/s) of this type of furnace, the intercritical annealing times were extended to 120 min (2 h) and 480 s (8 min) at 550 °C and 610 °C respectively, and this was followed by water quenching to ambient temperature. Henceforth, the samples annealed at 550 °C for 2 h are designated "550LA", and those annealed at 610 °C for 8 min are designated "610SA" where the LA and SA designations refer to the relatively long and short annealing treatments, respectively. Table 2 lists the phase ratios obtained from X-ray diffraction (XRD) analysis of both steel samples after intercritical annealing. It is seen that the final volume fractions of austenite are close to the intended $V_\gamma$ of 0.5. This is a key result and, as will be seen below, highlights the success of the present alloy design, since we are able to use two different intercritical annealing temperatures to achieve the same phase constitution, and yet one route is relatively rapid, and one is relatively slow. As will be demonstrated below, the fundamental nature of the austenite in each of these cases is fundamentally different.

## 3. Experimental Procedures

X–ray diffraction was performed on colloidal silica polished specimens using $Cu_{K\alpha}$ radiation. The peak shape analysis was performed using the TOPAS 3.0 software, fitted using the modified pseudo-Voigt function. The phase fractions were determined via Rietveld analysis. The dislocation density of the ferrite was evaluated using the approach detailed in [21, 22].

Tensile testing was conducted at a constant strain rate of 0.001 $s^{-1}$ using an MTS 810 mechanical testing system. The samples were machined into the ASTM E8 [23] sub-sized specimens (6 mm wide, and 25 mm gauge length) along the longitudinal direction of the annealed strips. To study the deformed state, samples were extracted from the gauge regions of the tensile samples. The strain of the each sample was estimated by the local area reduction.

Samples for TEM were mechanically polished using 1200 grit sand papers and then electropolished with an electrolyte of 10% perchloric acid and 90% acetic acid under the voltage of 15 V at a temperature of ∼10 °C. The thin foils were examined in a JEOL 2100 TEM, and a JEOL 2200FS TEM, both operating at 200 kV and well suited to the application of selected area electron diffraction (SAED) techniques These thin foils were also used for the SEM-TKD mapping in a Zeiss *Ultra* field-emission gun SEM operating at 30 kV. The experimental details of the novel SEM-TKD mapping technique, including the beam settings, optimisation procedures, and other relevant applications have recently been reviewed by Trimby and colleagues [24, 25]. The beam current was set to between 1–10 nA using the “high current” mode, and the mapping step was set to 4 nm. The TKD patterns were imaged using the Oxford Nordlys Nano

EBSD detector and the results were processed using the Oxford Instruments AZtec 2.0 EBSD software. The mapping results have a detection rate of above 70 %. This microscope was also equipped with an Oxford X-Max silicon drift X-ray detector for energy dispersive X-ray spectroscopy (EDXS) and this was used to map the macroscale distribution of mangense.

Specimens for atom probe microscopy (APM) were sampled from the annealed steels by cutting small blanks (~0.4 mm × 0.4 mm × 15 mm). Fine tips with a radius of 50–100 nm were prepared using the standard 2-stage electropolishing technique [26]. The APM experiments were performed using a local electrode atom probe (LEAP) 3000X Si with a voltage pulse repetition rate of 200 kHz, pulse fraction (the ratio of pulse voltage to DC standing voltage) of 20%, detection rate of 0.005 atoms per pulse, and specimen temperature of ~40 K. Reconstruction and visualisation of the APM dataset was performed using the Cameca's Image Visualization and Analysis Software (IVAS) package in combination with the advanced tomographic reconstruction approach by Gault et al. [27]. The LEAP instrument used in the current work had a detector efficiency of ~57% measured using the method described in the reference [28]. The detection of carbon ions in the mass spectra can result in peaks at 6, 12, 18, 24, and 36 m/z, and the carbon content is calculated by the approach recently discussed by Marceau et *al*. [29].

## 4 Experimental Results

### 3.1 Austenite-Ferrite Duplex Microstructure

Figures 2 and 3 provide the SEM-TKD crystallographic analyses for this novel steel in the 550LA and 610SA conditions, respectively. Specifically, these analyses include phase maps, pattern quality maps, and inverse pole figures in the Z direction, which is normal to the surface of the steel sheet. It is clear that the SEM-TKD maps provide a large field of view with nanoscale spatial resolution [24, 25]. Figures 2a and 3a demonstrate that both intercritically annealed samples possess similar microstructures, with approximately equal fractions of ferrite and austenite. Based on these SEM-TKD maps, the volume fractions and grain sizes of the ferrite and austenite phases were measured and the results summarised in Table 2. The confidence interval for the volume fractions was determined from the uncertainty in the area within the SEM-TKD map after noise reduction. Grain boundaries were defined as occurring where crystallographic misorientations exceeded 15 degrees, and grain size was measured using the mean lineal intercept method. The average ferrite grain size of 697±110 nm in the 610SA steel is larger than that of 612±91 nm in the 550LA steel. The average austenite grain sizes in both steels were just below 300 nm. Figures 2b and 3b reveal that there is a strong tendency for the {111} planes of the ferrite to align in the steel sheets. In other experimental results not shown here, the {100} planes of the ferrite also tend to align in the steel sheet. We propose that this ferritic texture is inherited from the cold-rolled martensite, which was subsequently intercritically annealed. The austenite exhibited less texture as is apparent from Figs 2c and 3c. This was attributed to the transformation crystallography during the martensite-to-austenite transformation. One austenitic grain can maintain an approximate Kurdjumov-Sachs (KS) orientation relationship (OR) [30] (viz.): $[1\ 0\ \bar{1}]_{\gamma}\ ||\ [1\ 1\ \bar{1}]_{\alpha}$, and $(1\ 1\ 1)_{\gamma}\ ||\ (0\ 1\ 1)_{\alpha}$ with the neighboring ferrite on one side as marked by the green lines in Figs 2 and 3. Since the austenite forms during the intercritical anneal via a diffusional transformation from the prior cold rolled martensite. As a consequence, it will not bear a good-fit orientation relationship with all the adjacent ferrite grains, thus weakening the inheritance of an austenitic texture.

Energy dispersive X-ray spectroscopy (EDXS) was conducted during the TKD mapping. However, the nature of the thin foil specimens is such that the intensity of the Mn Kα signal may be affected by the specimen thickness. Therefore, the (Mn Kα)/(Mn Kα +Fe Kα) Kα peak ratio of each mapped pixel was further computed so as to reveal clearly the distribution of maganese. In the 550LA steel, the manganese concentration in most austenite grains is displayed as blue, Fig. 2d. In contrast, the austenite in the 610SA steel contains less manganese, Fig. 3d, and the manganese is not distributed homogeneously, with lower levels occurring in the grain centres, consistent with a coring reaction. The intergranular variation in manganese enrichment in austenite in the 610SA steel was attributed to the short intercritical annealing time.

Figure 4a is a bright field (BF) TEM image of the duplex microstructure in the 550LA sample. The corresponding electron diffraction pattern (Fig. 4b) indicates that the α/γ orientation is close to the K-S OR [30]. It is also clear that dislocation debris from the cold-rolled martensite persists in the ferrite, subject to local variations. This indicates that the annealing at 550 °C for 2 h does not lead to complete recovery in the microstructure. Moreover, $V_4C_3$ carbides were be observed in the austenite grains as indicated in Fig. 4c. By contrast, the ferritic grains in the 610SA steel contain lower dislocation densities as revealed in Fig. 4d, and this is consistent with the X-Ray diffraction analysis in Table 2. Raising the annealing temperature to 610 °C clearly accelerates the recovery of dislocations. Very few $V_4C_3$ carbides were observed in ferrite or austenite of this sample. Hence, a greater yield strength might be expected in this 550LA steel because it contains more dislocations in the ferrite and the $V_4C_3$ precipitates occur in the austenite.

3.2 Mechanical Properties

Figure 5 provides the uniaxial engineering and true stress-strain curves for this steel after both intercritical annealing treatments. The 550LA treatment exhibits continuous yielding (Fig. 5a ) with a 0.2% proof strength of 1.24 GPa. However, the sample has limited ductility, with a tensile elongation of only 8.5%. By contrast, the 610SA treatment exhibits discontinuous yielding followed by steady work hardening. Although both heat treatments reach the GPa-level yield strengths, this subsequent work hardening in the 610SA condition enhances the ductility greatly such that the tensile elongation is over 25%, whilst the yield strength is 1.08 GPa, and the ultimate tensile strength is 1.39 GPa. These results are summarised in Table 3 together with other recent research results [31, 32] where similar properties have been achieved by enabling the formation of reverted austenite from nanoscale segregation in maraging-TRIP steels. As indicated in Table 3, impressive mechanical properties have been reported in other steels with manganese contents as low as 5 and 7 wt. % [2, 4, 15]. Nevertheless, the tensile properties obtained from this ductile steel in the 610SA condition are excellent, since both the yield and ultimate tensile strengths exceed 1 GPa. To our knowledge, this is the first time that such a new property has been reported in duplex steels. Furthermore, the annealing time is short enough to be implemented in a continuous heat-treatment production line.

To understand the enormous differences in the work hardening behaviour obtained for the two different intercritical annealing treatments, high-resolution experiments were carried out using atom probe microscopy.

3.3 Atom Probe Microscopy

Figure 6a provides tomographic atom maps of iron, manganese, vanadium and carbon across an austenite/ferrite interface in the 550LA condition. The grain enriched in manganese is austenite, and that which is depleted in manganese and other solutes is ferrite. APM also revealed the presence of a fine dispersion of nanoscale $V_4C_3$ carbides in the austenite, and these observations were supported by our TEM observations of fine-scale precipitation, such as in Fig. 4c. The [V]/[C] atom ratio in the carbide precipitate was approximately 1.38 ± 0.10, which is close to the ideal value of 1.33 for the chemical stoichiometry of $V_4C_3$. During the APM experiment, field desorption images were recorded. These record the individual ionic detector hit positions, and often reveal characteristic crystallographic information [26, 33]. In Fig. 6b, high density regions appear in red, and low density regions appear in green such that crystallographic zone lines and poles are clearly visible. Now, it is well knows that these regions represent something of an artefact in APT as the local point density of atoms is affected by the local crystallography of the specimen and, for example, C and N atoms have been known to directionally diffuse towards these locations due to the field gradients over the specimen surface [34]. Following the procedure of Stephenson et al. [35], we removed these volumes before computing the compositional profiles and other related chemical analyses in the current work. However, the crystallographic information revealed in the field desorption images is highly valuable for ensuring accurate tomographic reconstruction [27], and in the determination of iso-concentration surfaces. Using the fact that the lattice planes can be observed easily along the pole directions, and that crystal structural changes can be observed across the interfaces, we defined the interphase interface by adjusting the threshold to locate the iso-concentration surface at the precise point of transition in crystal structure. By tuning the 3D reconstruction map along the {011} pole, the {011} ferrite lattice planes become visible, and transition in crystal structure across the γ/α interface is clearly discernable, Fig. 6b. Thus, the threshold for the iso-concentration surface, which was set at 10 at. % manganese, was adjusted so as to locate the surface, at this transition. The compositional profiles of carbon, manganese, silicon, molybdenum, and vanadium across the iso-concentration surface were computed using a proxigram approach with a layer width of 0.25 nm. This approach [26, 36] computes the composition of each layer along the normal of the determined iso-concentration surface. The partitioning of carbon and manganese is significant, as revealed in Figs. 6c and d. In the austenite grain, the average manganese content was 16.3 at. %, and the average carbon content was 0.52 at. %, Fig. 6c-d. In the ferrite grain, the average manganese content was 2.9 at. %, and the average carbon content was 0.09 at. %. Significantly, these experiments have recorded striking compositional spikes of silicon and molybdenum in the vicinity of the austenite/ferrite interface.

Figures 7a-k provides the corresponding APM data for the steel in the 610SA condition. The tomographic atom maps for iron, manganese, vanadium and carbon are provided in Fig. 7a. Careful analysis of the field desorption images (inset) revealed that these tomographic atom maps include two γ/α interphase interfaces, separated by a γ/γ grain boundary. Although the crystallographic artefacts of austenite grains are dim due to the high solute content of this phase, the symmetry can still be identified, as revealed in Fig 7a. For example, the {111} pole in the first austenite grain ($\gamma_1$), is close to the {110} pole of the neighbouring ferrite grain ($\alpha_1$), which is consistent with the K-S OR.

Similarly, the second austenite grain ($\gamma_2$) exhibits a {100} pole close to the {110} pole of the neighbouring ferrite grain ($\alpha_2$).

Iso-concentration surfaces of 11 at. % and 8 at. % of manganese were created for the $\gamma_1/\alpha_1$ and $\gamma_2/\alpha_2$ interfaces, respectively as displayed in Fig. 7b. In the $\gamma_1$ austenite grain, the average manganese content was 15.0 at. %, and the average carbon content was 0.48 at. %, Fig. 7c-d. In the $\alpha_1$ austenite grain, the average manganese content was 3.4 at. %, and the average carbon content was 0.13 at. %. Careful inspection of these compositional profiles indicates some, albeit slight, enrichment of manganese or carbon at this interface. Much clearer evidence of manganese and carbon solute enrichment is available from the profiles across the $\gamma_2/\alpha_2$ interface, Fig. 7f-g. The averaged manganese content in that case is only 12.2 at. %, and the averaged carbon content is 0.44 at. % in the $\gamma_2$ austenite grain. Moreover, as shown in Fig. 7d and 7h, the partitioning of silicon and molybdenum is again not observed but their spikes at the interfaces are remarkable. However, the chemical spikes at the $\gamma_1/\gamma_2$ grain boundary are not seen in Fig. 7k. The origins of silicon and molybdenum spikes at the interface will be discussed in the section 4.1. The partitioning of vanadium is observed without spikes. In contrast to the 550LA steel, $V_4C_3$ carbides were seldom observed via TEM or APM in the 610SA steel. The average solute contents listed in Table 4 for austenite and ferrite grains were based on three separate APM experiments. These high-resolution chemical analysis data was then applied to understand the austenite stability and its deformation behaviour.

### 3.4 Deformation-Induced Microstructure

Figure 8 shows the evolution of the deformed microstructure as strain increased in the 550LA steel. The sequence of TEM images was recorded for different levels of reduction in area, Ar, during the tensile experiment. As shown in Figs 8a-b, the ferrite absorbs the imposed strain by dislocation slip, whereas the austenite absorbed the imposed strain by dislocation slip and faulting. There are intense dislocation tangles in the ferrite, and there are many faults as well as dislocations in the austenite. These faults appear to nucleate at the austenite grain boundaries or interfaces, as illustrated in the sub-panel of Fig. 8b. They are responsible for the latter formation of deformation-induced ε martensite or twinning. The fine structures of the deformation faults were not characterized here, but natures of the faults for ε martensite and twinning are different [37]. As the steel was strained to a 6.8 % reduction in area (Fig. 8c), extensive ε martensite and mechanical twins developed from the faults in the austenite. The corresponding SAED pattern in Fig. 8d reveals the twinning relationship and confirms the orientation relationship between austenite and ε martensite: $[2\ \bar{1}\ \bar{1}\ 0]_{\varepsilon}\ ||\ [1\ 0\ \bar{1}]_{\gamma}$, and $(0\ 0\ 0\ 1)_{\varepsilon}\ ||\ (1\ 1\ 1)_{\gamma}$. It was clear that the formation of α' martensite within the austenite was not observed in the 550LA steel, even though this austenite composition is located in the ε+α' transformation field in Schumann's austenite stability map for Fe-Mn-C alloys [38]. In fact, our results demonstrate that the primary deformation mechanism of austenite here is dislocation slip together with the formation of ε martensite and mechanical twinning. It is noted there are many austenitic grains containing a lamellar fault structure where only a single fault variant operates, Fig. 8c. In fact, multiple fault variants were seen to operate in the larger austenitic grains [2, 10], but only a single variant of the faults was observed in smaller grains, Fig 8b-c. Based on TEM results, the single lamella structure can be observed in many austenite grains, especially when grain size is below about 300 nm. This is in sharp contrast to the observations in other duplex steels with coarser austenite grains [2, 10].

The evolution of the deformation microstructure for the 610SA steel is set out in Fig. 9 and during the initial stages, preceded similarly to that for the 550LA steel. The yielding in ferrite is via dislocation glide, and the deformation-induced defects in austenite involve dislocations and faulting, but mechanical twinning hasn't been observed. When the strain level reached 5.5 %, α' martensite, together with ε martensite, was observed in austenite, as shown in Fig. 9a,b. Most of the α' martensite plates were observed to be in contact with austenite grain boundaries, indicating that these were the favoured nucleation sites. The approximate three-phase crystallography could be established from the SAED patterns as provided in Fig. 9c: $[1\ 0\ 0]_{\alpha'} \| [2\ \bar{1}\ \bar{1}\ 0]_{\varepsilon} \| [1\ 0\ \bar{1}]_{\gamma}$, and $(0\ 1\ 1)_{\alpha'} \| (0\ 0\ 0\ 1)_{\varepsilon} \| (1\ 1\ 1)_{\gamma}$. The HRTEM image in Fig. 9d reveals the transformation front of the deformation-induced α' martensite (arrowed). Significantly, α' martensite could penetrate into pre-existing bands of ε martensite. As shown in Figs. 9e and 9f, α' martensite eventually occupied the whole austenite grain and took the form of twinned martensitic laths when the strain levels were further increased. Hence, the deformation mechanism of austenite follows the sequence: ε→ α'.

The results of the phase analysis by XRD and SEM-TKD mapping near the fracture surface are set out in Table 2. The SEM-TKD maps near the fracture surfaces of two annealed steels are provided in Fig. 11. It is clear that the crucial difference in the two conditions is the capacity to trigger the deformation-induced α' martensite from austenite.

## 4 Discussions

4.1 Solute partitioning during the phase transformation

Figure 10a summarises the solute contents from the APM experiments (Table 4) and the calculated solute contents in austenite and ferrite with respect to temperature at equilibrium. The carbon and manganese contents in austenite and ferrite in the steel annealed at 550 °C for 2 h are more close to the values calculated at equilibrium. The measured carbon and manganese contents in the steel annealed at 610 °C for 8 min deviate more from the equilibrium values. Three factors, all of which are attributed to the short annealing time, are now discussed in detail. Firstly, when the annealing time is short, austenite grains that possess a range of manganese concentrations will form. For example, the manganese enrichments in $\gamma_1$ and $\gamma_2$ in Fig. 7 are very different. This is supported by the Mn $K_\alpha$ mapping in Fig. 2d and 3d. Indeed, intragranular gradients and intergranular differences in the manganese contents must exist when the annealing time is short. However, these differences cannot explain the carbon contents in austenite exceeding the equilibrium values at 610 °C, Fig. 10a. Also, the $\gamma_1$ austenite grain in Fig. 7 has a higher carbon content than the value predicted (0.36 at. %) at 610 °C. Here, a second factor must be considered (viz.) that the initial solute partitioning does not match the equilibrium condition because $V_4C_3$ carbides cannot form in such a short time. Moreover, there were no $Mo_2C$ carbides observed in the samples after two annealing treatments. Figure 10b demonstrates solute contents in austenite and ferrite with respect to temperature based on a constrained computation by suspending the $Mo_2C$ and $V_4C_3$ phases. This calculation allows higher carbon contents to partition into austenite than is allowed under equilibrium conditions. The experimental carbon contents in austenite are closer to the predicted values (0.47 at. %) in Fig. 10b. Moreover, the vanadium contents in ferrite and austenite in the 610SA steel are also close to the results of the constrained computation. Hence, the fact that carbide precipitation is not significant at the initial transformation at 610 °C means that the solute partitioning is not completely as per that at the equilibrium. The third and final factor that we consider is the fact that the carbon segregation or even cementite precipitation formed during heating could be retained after such a short annealing time. This explains the very high carbon accumulation (~ 0.88 at. %) at the interface in Fig. 7d and Fig. 10. Such nanoscale segregation was also used in the design of other duplex steels [32]. The solute partitioning in the 610SA steel can be affected by all these factors whereas their effects can be ignored in the case of the longer annealing times such as was demonstrated in the 550LA steel.

The compositional spikes at the interface can originate from non-partition local equilibrium [39] or interfacial segregation of solute [40]. Figures 6 and 7 reveal that very limited silicon and molybdenum partitioning occurs between ferrite and austenite after the annealing treatments. Based on the thermodynamic calculation results provided in Fig. 10, austenite should be enriched with silicon and ferrite should be enriched with molybdenum, although the partitioning levels are not expected to be high. However, the observed solute spikes at the interface are very high and they greatly exceed the solubilities in austenite and ferrite at equilibrium. The local equilibrium (negligible partitioning or partitioning) condition does not apply because it requires that all solutes have equal chemical potential in all phases that are in contact at the interface. Hence, it is suggested that the observed silicon and molybdenum spikes originate from interfacial segregation. Finally, we note that this nanoscale interfacial segregation cannot significantly affect the stability of the austenite because their compositional levels are not high enough to suppress the martensitic transformation.

4.2 Stability of Ultrafine-Grained Austenite

The essential concept of austenite engineering is to tailor the manganese and carbon contents in the austenite so as to obtain metastable austenite after annealing, and to thus trigger martensitic TRIP reactions to deformation. Therefore, the martensite transformation start (*Ms*) temperature may be regarded as parameter that directly controls the stability of the austenite. Using the APM results summarised in Table 3, the *Ms* temperatures of the α' martensite and ε martensite can be estimated by the following empirical equations:

$M_s\ (°C) = 539 - 423[\mathrm{C}] - 30.4[\mathrm{Mn}] - 7.5[\mathrm{Mo}]$ for α’-martensite [41], and (1)

$M_s\ (°C) = 303 \pm 8 - (489 \pm 31)[\mathrm{C}] + (4.1 \pm 1)[\mathrm{Si}] - (9.1 \pm 0.4)[\mathrm{Mn}] - (19.4 \pm 5)[\mathrm{Mo}] - (34 \pm 10)[\mathrm{V}]$ for ε martensite [42], (2)

where [X] is the chemical composition of element X in wt. %. The *M*s temperatures (designated as $Ms^0$) based purely on the chemical compositions of austenite were estimated by Eq. (1) and (2) . As listed in Table 5, the estimated $Ms^0$ temperatures are very high and are inconsistent with the TEM observations and the measurements from both the dilatometer and differential scanning calorimeter. We propose that this discrepancy occurs because of the effects of austenite grain size on the martensitic transformation. Following on from the ideal Koistinen-Marburger equation, the relationship between austenite grain size and *M*s temperature was further derived by Yang and Bhadeshia [43] as:

$$M_s^0 - T = \frac{1}{b}\ln\left[\frac{1}{V_\gamma}\left\{exp\left(\frac{\ln(1-f_M)}{m}\right) - 1\right\} + 1\right] \quad , \quad (3)$$

where $f_M$ is the volume fraction of martensite, $V_\gamma$ is the average grain volume of austenite, $M_s^0$ is the *Ms* temperature when $V_\gamma$ is very large, *T* is the temperature at which $f_M$ volume fraction of martensite is measured, *m* is the aspect ratio of the martensite plate (0.05 for α-martensite and 0.03 for ε-martensite). The constant *b* is 0.2689 for $\alpha$' martensite based on [43], while it is proposed to be 0.19 for $\varepsilon$ martensite to fit to the measured data in Table 5. The values of $Ms^0$ were calculated using Eq. 3, and the *Ms* temperatures that account for the austenite grain size (designated as $Ms^g$) were taken for $f_M$ = 0.01. Hence, the experimental and theoretical results show that the effects of austenite grain size in suppressing the *Ms* temperatures of α' martensite and ε-martensite are, in fact, remarkable. The estimated $Ms^g$ temperature with the effect of grain size is consistent with the experimental results. The analysis indicates clearly that the austenite in the 610SA steel is less stable than that in the 550LA steel, which is consistent with the fact that more α' martensite was found in the former after fracture. Moreover, due to the size effect on the *Ms* temperatures, austenitic grains, which ought to have transformed into martensite, can be preserved at room temperature. This leads to the success of both annealing conditions to create ultrafine duplex microstructure. The annealing at 610 $^{o}$C for 8 min is especially promising because it also achieves an excellent mechanical property.

4.3 On the Effect of Austenite Stability on Deformation-Induced Microstructure

One consequence of the high level of grain refinement was the efficacy of grain boundaries or α/γ interfaces as nucleation site for faults. This differs from the observations related to intragranular nucleation models for ε martensite or twinning, such as the pile-up model [44, 45], the stair-rod model [46], the spiral model [47] and the super-stacking-fault model [48]. Based on our TEM investigations, the ε martensite and twinning primarily nucleate at grain boundaries or interfaces as demonstrated in Fig 8b. The emission of faults from grain boundaries has also been noted in ultrafine-grained and nanocrystalline 301LN stainless steels [49]. A related finding from this work is that the lamellar fault structure is prevalent when the austenite grain sizes are small. This is consistent with the observation that significant grain refinement can change a multiple twin structure to a single variant lamellar twin structure in Fe-22Mn-0.6C (wt. %) [50]. We propose that the fault formation from these boundaries and interfaces is a stress-controlled process. A higher local stress will be required to trigger other fault systems, because of the frequent need to propagate across pre-existing lamellar faults, ε-martensite or twin structures due to the constrained volume.

The dominant nucleation site for deformation-induced α' martensite was the austenite grain boundary in ultrafine-grained austenite. Indeed, it is known that α' martensite prefers to nucleate at grain boundaries [39]. Nucleation of the α' martensite phase at grain boundaries during deformation is commonly reported to be a stress-controlled or stress-assisted nucleation process [51]. This is important in explaining the observed TRIP effects here, since it has long been suggested that the intersections of shear bands should serve as the predominant sites for the nucleation of strain-induced martensite [52]. The principle is that the shear strains by two variants of ε martensite lead to an atomic arrangement favourable for α' martensite nucleation [44, 53, 54] as shown in Fig. 11. However, Das et al. [55] recently found that the ratio of α' martensite to plastic strain can be described in terms of the thermodynamic effect of applied stress.

In the 610SA steel, α' martensite normally nucleates from the austenite boundary as shown in Fig. 9b. This is primarily attributed to the grain size refinement. As suggested above, the multiple-variant fault structure is difficult to form when the grain size is small. As shown in Fig. 11, we suggest that, when the austenite is reduced to a fine scale of about 300 nm, the primarily initial deformation-induced microstructure is the single lamella structure of ε martensite or mechanical twinning, which cannot provide sufficient nucleation sites for strain-induced martensite. Hence, stress-assisted martensite becomes a relative important mechanism in this work.

Moreover, our experimental finds that the compositional differences of only 2.6 wt. % of manganese in austenite between two annealed steels is a significant factor in the efficiency of the TRIP effect. We now explain this suggestion in more details based on the principles of stress-assisted martensitic transformation. The thermodynamics model of stress-assisted martensitic transformation can therefore be expressed as:

$$\Delta G_{\mathrm{CHEM}} + \Delta G_{\mathrm{MECH}} > E_{\mathrm{STORE}}. \tag{4}$$

The chemical driving force for transformation $\Delta G_{\mathrm{CHEM}} = G_{\gamma} - G_{\alpha}$ (1891 J.mol$^{-1}$ for the 550LA steel and 2396 J.mol$^{-1}$ for the 610SA steel) was calculated using Thermo-Calc with the TCFE7 database at 27 °C, using the chemical compositions listed in Table 4. The mechanical driving force is given by $\Delta G_{\mathrm{MECH}} = 0.86\sigma$, where $\sigma$ is the external stress [56]. $E_{\mathrm{STORE}}$ is the stored energy of the martensite. The grain size effect on the stability of the austenite in the two steels is proposed to be the same. The chemical driving forces in the 550LA steel is less than that in the 610SA steel by ~505 J.mol$^{-1}$. Hence, an additional stress of 587 MPa is required to assist the deformation-induced α' martensite in the 550LA steel. Assuming α' martensite transformation could be triggered at the true stress of 1200 MPa (true strain is about 4 %) in the 610SA steel, the 550LA steel will require a stress of approximately 1787 MPa to activate the stress-assisted α' martensite transformation. This explains why α' martensite in the 550LA steel does not efficiently occur during deformation. The thermodynamics model proves that the stress-assisted process for the deformation-induced α' martensite transformation is very sensitive to the chemical composition of austenite, and that the Mn concentration is of critical importance. Moreover, the TRIP behaviour via a stress-assisted process at the grain boundary is especially important in the duplex steels with fine grain size of this work. This is different from the strain-induced martensite formation that operates in other manganese-rich duplex steels with larger austenite grains size [2, 4, 10, 15].

4.4 Work Hardening Rate in Ultrafine-Grained Duplex Steels

As summarized in Fig. 12, the TRIP effect via ε martensite and α' martensite formation from the deformed austenite in the 610SA steel contributes a substantial capacity for work hardening in contrast to the TRIP effect via ε martensite coupled with the a minor TWIP effect in the 550LA steel. On the other hand, the chemical effect on austenite stability makes the austenite in the 550LA steel much more stable, since it is relatively resistant to the formation of ε martensite and twinning. The other half of the microstructure, (viz.) the ferrite, is so fine that the strain softening occurs very quickly [8]. The work hardening rate in the 550LA steel dropped below the true stress after a strain of about 7% as shown in Fig. 12. Although ε martensite, deformation twins, and even nanometre-sized carbides can enhance the strain hardening rate, their contributions are insufficient to prevent the 550LA steel from overcoming plastic instability. As is evident in Fig. 12, the 610SA steel has additional capacity for strain hardening due to the deformation-induced α' martensite. These observations are consistent with reports that α' martensite can contribute a higher work hardening than mechanical twinning and ε martensite [57]. Hence, the 550LA austenite is very stable, and reaches the plastic instability, $\frac{\partial\sigma}{\partial\varepsilon} < \sigma$, before the stress threshold required for the onset of the α' martensite transformation is reached in most austenite grains. The higher dislocation density in ferrite and the nanometre carbides in austenite enhance the initial work hardening rate of the 550LA steel as shown in Fig. 12 but they cannot assist a sustainable work hardening for the latter deformation. Therefore, austenite engineering such as proposed here, via careful selection of alloy composition and thermomechanical treatment, enables control of the onset of deformation-induced α' martensite which in turn achieves excellent ductility and GPa-level yield strength. This work demonstrates that the stability of the austenite needs to be marginal in order to ensure a capacity for work hardening rate, and hence the choice of the higher annealing temperature.

## 5 Conclusions

Based on microstructural observations from TKD-mapping, TEM, and especially APM, this work shows that chemically instable austenite can be preserved as stable phase at room temperature when a significant reduction in grain size is achieved. We validated this idea by annealing at 610 °C for 8 min and at 550 °C for 2 h respectively to create ultrafine-grained 50/50 austenite-ferrite duplex microstructure in a newly-designed alloy. This idea enables annealing at higher temperatures for short duration so as to create new options in time-temperature space for duplex steels.

When the new alloy was subjected to a short annealing at 610 °C, YS of 1080 MPa, UTS of 1390 MPa, and a total elongation of 26.3 % were achieved. Alternatively, a longer annealing at 550 °C resulted in YS of 1240 MPa, UTS of 1360 MPa, and a total elongation of only 8.5 %. The enhanced work hardening from the deformation-induced α' martensite is significant, when compared with the contributions to work hardening from deformation-induced ε martensite, mechanical twinning, or even nanoscale carbides. Significantly, we concluded that a high volume fraction of austenite does not always impart ductility for ultrafine-grained duplex steels.

We present experimentally and theoretically evidence to assert the proposal that the compositional differences as small as 2.6 wt. % of manganese greatly influence the deformation-induced microstructure and mechanical properties. This suggests here that the austenite chemistry must be carefully controlled in order to obtain the desired TRIP effect. In particular, when grain size is small (~300 nm), the initial deformation-induced ε martensite and twinning in many grains will degenerate into the operation of a single variant, resulting in a lamella microstructure. This is unfavourable for the nucleation of strain-induced martensite. The TEM evidence suggest that α'-TRIP effect in this situation is enhanced predominately by the stress-assisted process, rather than the more commonly discussed strain-induced nucleation. Hence, in the regime of ultrafine-grained austenite, special attention should be paid to the chemical composition of austenite, since this greatly impacts the onset of stress-assisted martensite. These results are shown to be consistent with the analysis based on the thermodynamics model of stress-induced martensite. Recent work [58] has suggested very small austenite grains may in some circumstances be less stable to mechanically-induced transformation than large ones. This is because the deformation-induced transformation in larger regions of austenite is delayed by prior mechanical twinning which is less prevalent in the smaller austenite regions. The present work did not reveal such behavior, possible because when twinning occurred, it was on single variants and was preceded by significant dislocation plasticity.

**Acknowledgement**

The authors acknowledge the facilities, and the scientific and technical assistance of the Australian Microscopy & Microanalysis Research Facility (ammrf.org.au) node at Sydney Microscopy & Microanalysis, at the University of Sydney. The authors are particularly grateful to Drs. Takanori Sato and Patrick Trimby for their assistance in atom probe microscopy and transmission Kikuchi diffraction experiments, respectively. We acknowledge gratefully Professors Dierk Raabe and Dirk Ponge at the Max-Planck-Institut für Eisenforschung, GmbH for providing the results of their research for the purpose of the comparison in Table 3. The authors are grateful to the Australian Research Council for partial support of this work.

**References**

[1] Bouaziz O, Zurob H, Huang M. Steel Research International 2013;84:937.

[2] De Cooman BC, Gibbs P, Lee S, Matlock DK. Metallurgical and Materials Transactions a-Physical Metallurgy and Materials Science 2013;44A:2563.

[3] Edmonds DV, He K, Rizzo FC, De Cooman BC, Matlock DK, Speer JG. Materials Science and Engineering a-Structural Materials Properties Microstructure and Processing 2006;438:25.

[4] Luo H, Shi J, Wang C, Cao W, Sun X, Dong H. Acta Materialia 2011;59:4002.

[5] Ma E. Scripta Materialia 2003;49:663.

[6] Meyers MA, Mishra A, Benson DJ. Progress in Materials Science 2006;51:427.

[7] Okitsu Y, Takata N, Tsuji N. Scripta Materialia 2011;64:896.

[8] Tsuji N, Ito Y, Saito Y, Minamino Y. Scripta Materialia 2002;47:893.

[9] Son YI, Lee YK, Park K-T, Lee CS, Shin DH. Acta Materialia 2005;53:3125.

[10] Herrera C, Ponge D, Raabe D. Acta Materialia 2011;59:4653.

[11] Yuan L, Ponge D, Wittig J, Choi P, Jimenez JA, Raabe D. Acta Materialia 2012;60:2790.

[12] Bouaziz O, Allain S, Scott CP, Cugy P, Barbier D. Current Opinion in Solid State & Materials Science 2011;15:141.

[13] Avishan B, Garcia-Mateo C, Morales-Rivas L, Yazdani S, Caballero FG. Journal of Materials Science 2013;48:6121.

[14] Bhadeshia HKDH. Proceedings of the Royal Society a-Mathematical Physical and Engineering Sciences 2010;466:3.

[15] Suh DW, Ryu JH, Joo MS, Yang HS, Lee K, Bhadeshia HKDH. Metallurgical and Materials Transactions a-Physical Metallurgy and Materials Science 2013;44A:286.

[16] Lee S, Lee S-J, De Cooman BC. Scripta Materialia 2011;65:225.

[17] Luo H. Scripta Materialia 2012;66:829.

[18] Andersson JO, Helander T, Hoglund LH, Shi PF, Sundman B. Calphad-Computer Coupling of Phase Diagrams and Thermochemistry 2002;26:273.

[19] Lee S-J, Lee S, De Cooman BC. International Journal of Materials Research 2013;104:423.

[20] Takahashi M, Bhadeshia HKD. Journal of Materials Science Letters 1989;8:477.

[21] Williamson GK, Smallman RE. Philosophical Magazine 1956;1:34.

[22] Williamson GK, Hall WH. Acta Metallurgica 1953;1:22.

[23] ASTM Standard E8/E8M-2013, Standard Test Methods for Tension Testing of

Metallic Materials. West Conshohocken, PA: ASTM International, 2013.

[24] Trimby PW. Ultramicroscopy 2012;120:16.

[25] Trimby PW, Cao Y, Chen Z, Han S, Hemker KJ, Lian J, Liao X, Rottmann P, Samudrala S, Sun J, Wang JT, Wheeler J, Cairney JM. Acta Materialia 2014;62:69.

[26] Gault B, Moody MP, Cairney JM, Ringer SP. Atom Probe Microscopy. New York: Springer, 2012.

[27] Gault B, Moody MP, de Geuser F, Tsafnat G, La Fontaine A, Stephenson LT, Haley D, Ringer SP. Journal of Applied Physics 2009;105.

[28] Gault B, de Geuser F, Stephenson LT, Moody MP, Muddle BC, Ringer SP. Microscopy and Microanalysis 2008;14:296.

[29] Marceau RKW, Choi P, Raabe D. Ultramicroscopy 2013;132:239.

[30] Kurdjumov GV, Sachs G. Z. Phys. 1930;64:325.

[31] Raabe D, Ponge D, Dmitrieva O, Sander B. Scripta Materialia 2009;60:1141.

[32] Raabe D, Sandloebes S, Millan J, Ponge D, Assadi H, Herbig M, Choi PP. Acta Materialia 2013;61:6132.

[33] Yao L, Moody MP, Cairney JM, Haley D, Ceguerra AV, Zhu C, Ringer SP. Ultramicroscopy 2011;111:458.

[34] Gault B, Danoix F, Hoummada K, Mangelinck D, Leitner H. Ultramicroscopy 2012;113:182.

[35] Stephenson LT, Moody MP, Liddicoat PV, Ringer SP. Microscopy and Microanalysis 2007;13:448.

[36] Felfer PJ, Gault B, Sha G, Stephenson L, Ringer SP, Cairney JM. Microscopy and Microanalysis 2012;18:359.

[37] Brooks JW, Loretto MH, Smallman RE. Acta Metallurgica 1979;27:1829.

[38] Schumann VH. Neue Hütte 1971;17:605.

[39] Honeycombe RKW, Bhadeshia HKDH. Steels: Microstructure and Properties, 3rd: Elsevier Ltd. International, 2006.

[40] Gupta D. MTA 1977;8:1431.

[41] Andrews KW. Journal of the Iron and Steel Institute 1965;203:721.

[42] Yang HS, Jang JH, Bhadeshia HKDH, Suh DW. Calphad-Computer Coupling of Phase Diagrams and Thermochemistry 2012;36:16.

[43] Yang HS, Bhadeshia HKDH. Scripta Materialia 2009;60:493.

[44] Lecroisey F, Pineau A. Metallurgical Transactions 1972;3:391.

[45] Cohen JB, Weertman J. Acta Metallurgica 1963;11:996.

[46] Fujita H, Ueda S. Acta Metallurgica 1972;20:759.

[47] Venables JA. Philosophical Magazine 1974;30:1165.

[48] Mahajan S, Green ML, Brasen D. MTA 1977;8:283.

[49] Misra RDK, Kumar BR, Somani M, Karjalainen P. Scripta Materialia 2008;59:79.

[50] Gutierrez-Urrutia I, Raabe D. Scripta Materialia 2012;66:992.

[51] Olson GB, Cohen M. Annual Review of Materials Science 1981;11:1.

[52] Olson GB, Cohen M. Journal of the Less-Common Metals 1972;28:107.

[53] Suzuki T, Kojima H, Suzuki K, Hashimoto T, Ichihara M. Acta Metallurgica 1977;25:1151.

[54] Lee T-H, Ha H-Y, Kang J-Y, Moon J, Lee C-H, Park S-J. Acta Materialia 2013;61:7399.

[55] Das A, Chakraborti PC, Tarafder S, Bhadeshia H. Materials Science and Technology 2011;27:366.

[56] Olson GB, Cohen M. MTA 1982;13:1907.

[57] Frommeyer G, Brux U, Neumann P. Isij International 2003;43:438.

[58] Wang MM, Tasan CC, Ponge D, Kostka A, Raabe D. Acta Materialia 2014;79:268.

**Tables & Table Captions**

Table 1 The chemical composition of the studied steel

| Fe | C | Mn | Si | P | S | V | Mo | N | |
|---|---|---|---|---|---|---|---|---|---|
| Bal. | 0.08 | 10.60 | 0.36 | 0.01 | 0.01 | 0.07 | 0.28 | 0.003 | in wt. % |
| Bal. | 0.37 | 10.70 | 0.68 | 0.02 | 0.02 | 0.08 | 0.16 | 0.01 | in at. % |

Table 2 The phase volume fractions and the grain sizes in the duplex steels

| X-Ray Diffraction Analysis | | | | |
|---|---|---|---|---|
| Condition | Vγ | Vα | dislocation density in α ($m^{-2}$) | |
| 550LA | 0.45±0.01 | 0.55±0.01 | $5.8 \times 10^{13}$ | |
| 610SA | 0.54±0.01 | 0.46±0.01 | $1.6 \times 10^{13}$ | |
| 550LA-deformed* | 0.39±0.01 | 0.61±0.01 | $3.7 \times 10^{14}$ | |
| 610SA-deformed* | 0.07±0.01 | 0.93±0.01 | $1.1 \times 10^{14}$ | |
| SEM-TKD Mapping Analysis | | | | |
| Condition | Vγ | Vα | dγ **(nm)** | dα **(nm)** |
| 550LA | 0.48±0.04 | 0.52±0.04 | 288±35 | 612±91 |
| 610SA | 0.53±0.05 | 0.47±0.05 | 298±49 | 697±110 |
| 550LA-deformed* | 0.39±0.05 | 0.57±0.05 | - | - |
| 610SA-deformed* | 0.10±0.08 | 0.90±0.08 | - | - |

*measured at the position about 1.5 mm from the fracture surface

Table 3 Mechanical properties of Mn-based duplex steels

| Material | YS (MPa) | UTS (MPa) | EL (%) | Ref. |
|---|---|---|---|---|
| 11 Mn duplex steel, 550 $^{o}$C/2 h | 1240 | 1360 | 8.5 | a |
| 11 Mn duplex steel, 610 $^{o}$C/8 min | 1080 | 1390 | 26.7 | a |
| 9 Mn TRIP-Maraging, 450 $^{o}$C/48 h | 984 | 1015 | 15.3 | [36] |
| 9 Mn TRIP-Maraging, 450 $^{o}$C/48 h | 978 | 1011 | 12.7 | [36] [b] |
| 12 Mn TRIP-Maraging, 450 $^{o}$C/48 h | 1142 | 1322 | 21.6 | [36] |
| 12 Mn TRIP-Maraging, 450 $^{o}$C/48 h | 1394 | 1465 | 10.3 | [36] [b] |
| 5Mn duplex steel, 650 $^{o}$C/144 h | ~500 | ~960 | ~45 | [11] |
| 5Mn duplex steel, 650 $^{o}$C/1 min | ~830 | ~960 | ~20 | [11] |
| 7Mn duplex steel, 600 $^{o}$C/168 h | 700 | N/A | ~32 | [9] [b] |
| 5Mn2Al duplex steel, 700 $^{o}$C/260 s | ~750 | ~850 | ~35 | [20] |
| 5Mn2Al duplex steel, 760 $^{o}$C/272 s | ~650 | ~1000 | ~20 | [20] |

[a] this work, where the materials were cold rolled before annealing.
[b] published work, where materials were cold rolled before annealing.

Table 4 The solute contents in austenite and ferrite measured by the APM

| Condition | Phase | Fe | C | Mn | Si | V | Mo |
|---|---|---|---|---|---|---|---|
| 550LA | α | Bal. | 0.09 | 3.20 | 0.72 | 0.07 | 0.17 |
| (in at. %) | γ | Bal. | 0.49 | 16.52 | 0.66 | 0.03 | 0.16 |
| | α | Bal. | 0.02 | 3.16 | 0.36 | 0.06 | 0.29 |
| (in wt. %) | γ | Bal. | 0.11 | 16.40 | 0.33 | 0.03 | 0.28 |
| 610SA | α | Bal. | 0.15 | 4.02 | 0.74 | 0.14 | 0.17 |
| (in at. %) | γ | Bal. | 0.45 | 13.90 | 0.72 | 0.04 | 0.17 |
| | α | Bal. | 0.03 | 3.98 | 0.37 | 0.13 | 0.29 |
| (in wt. %) | γ | Bal. | 0.10 | 13.80 | 0.37 | 0.04 | 0.29 |

Table 5 The $Ms^0$, $Ms^g$, and measured *Ms* temperatures (in $^oC$)

| Sample | α-$Ms^0$ | ε-$Ms^0$ | dγ (μm) | α-$Ms^g$ | ε-$Ms^g$ | measured *Ms* |
|---|---|---|---|---|---|---|
| Test alloy[a] 1200 $^oC$ for 3 min | (68) | 143 | 300 | (63) | 133 | 138 (ε)[b] |
| Test alloy[a] 850 $^oC$ for 5 min | (68) | 143 | 10 | (26) | 79 | 82 (ε)[b] |
| Current alloy 850 $^oC$ for 5 min | 181 | (158) | 6 | 132 | (86) | 121 (α')[b] |
| 550LA | (-8) | 95 | 0.29 | (-90) | -25 | -25.5 (ε)[c] |
| 610SA | (75) | 123 | 0.30 | (-8) | 4 | 2.9 (ε+α')[c] |

[a]Fe-0.036C-15.2Mn-0.27Si-0.28Mo-0.01S-0.01P (in wt. %)
[b]measured by the dilatometer
[c]measured by the differential scanning calorimeter

**Figures & Figure Captions**

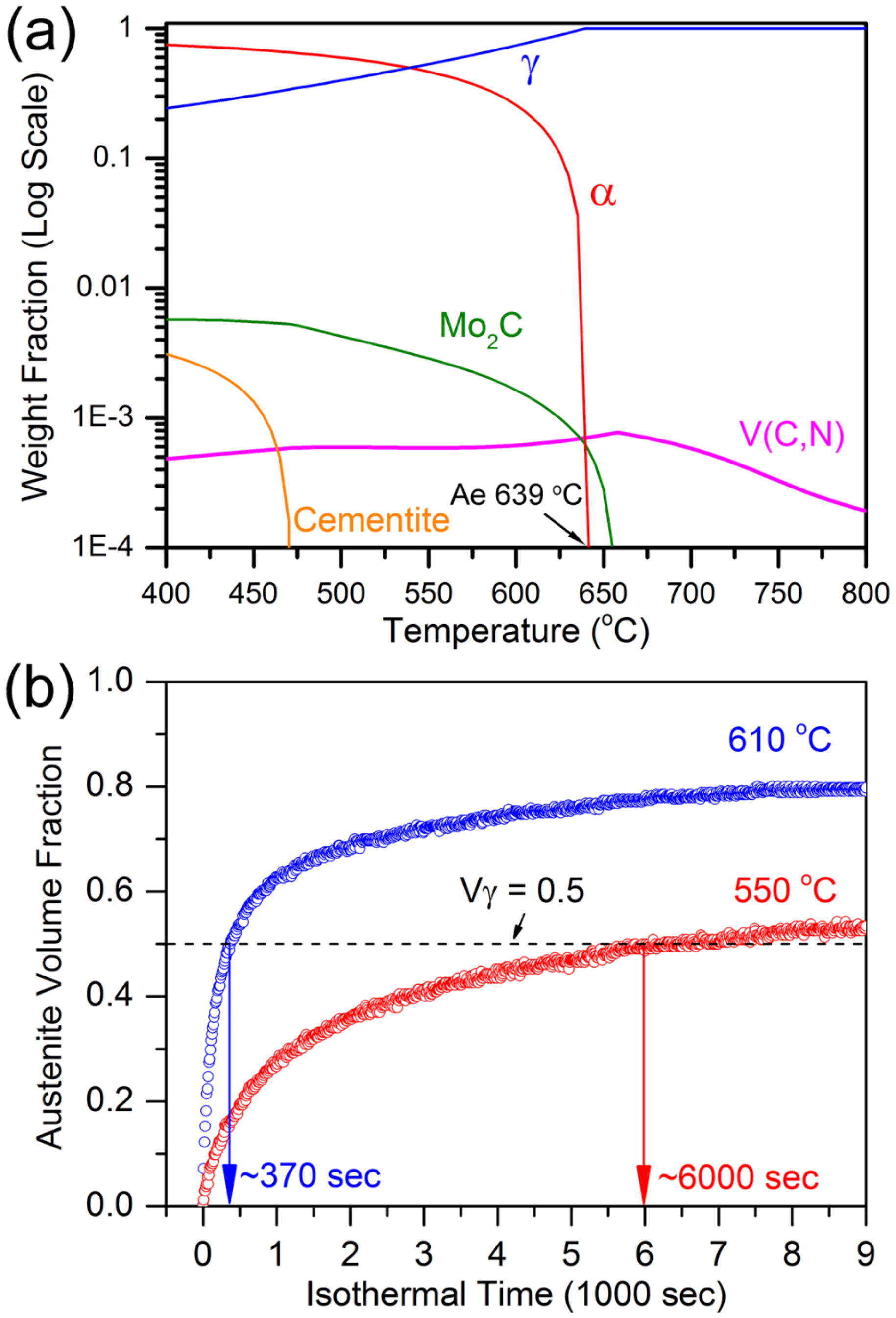

Figure 1 (a) The equilibrium phase fractions of the alloys at different temperature, computed using Thermo-Calc with the TCFE7 database, and (b) the volume fractions of austenite as a function of isothermal transformation time at 610 °C and 550 °C, measured using dilatometry.

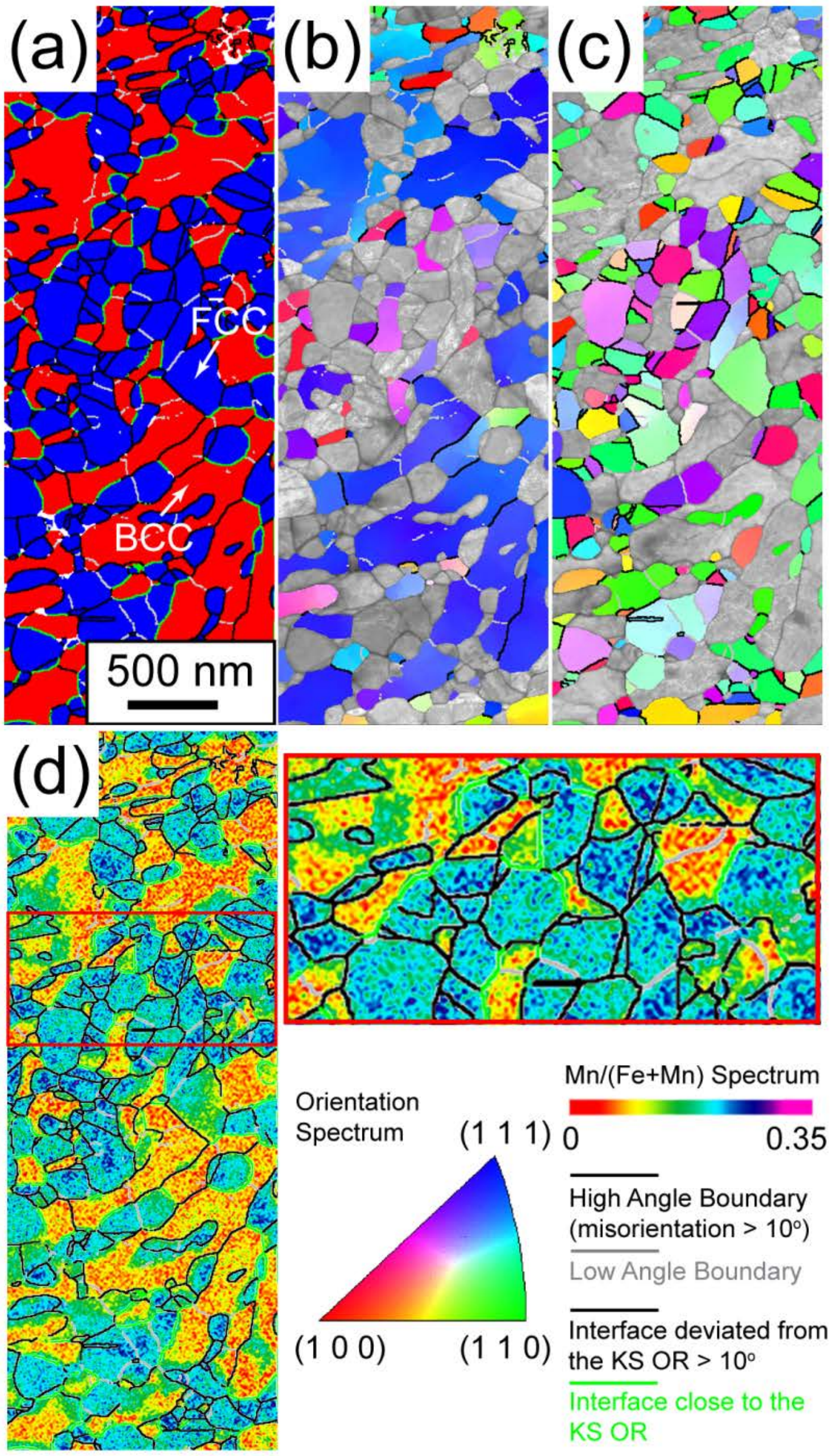

Figure 2 The SEM-TKD analyses on the 550LA steel: (a) the phase map showing ferrite (red) and austenite (blue), (b) the inverse pole figure-Z of ferrite, (c) the inverse pole figure-Z of austenite, (d) the EDXS Mn Kα /(Mn Kα +Fe Kα) peak ratio mapping.

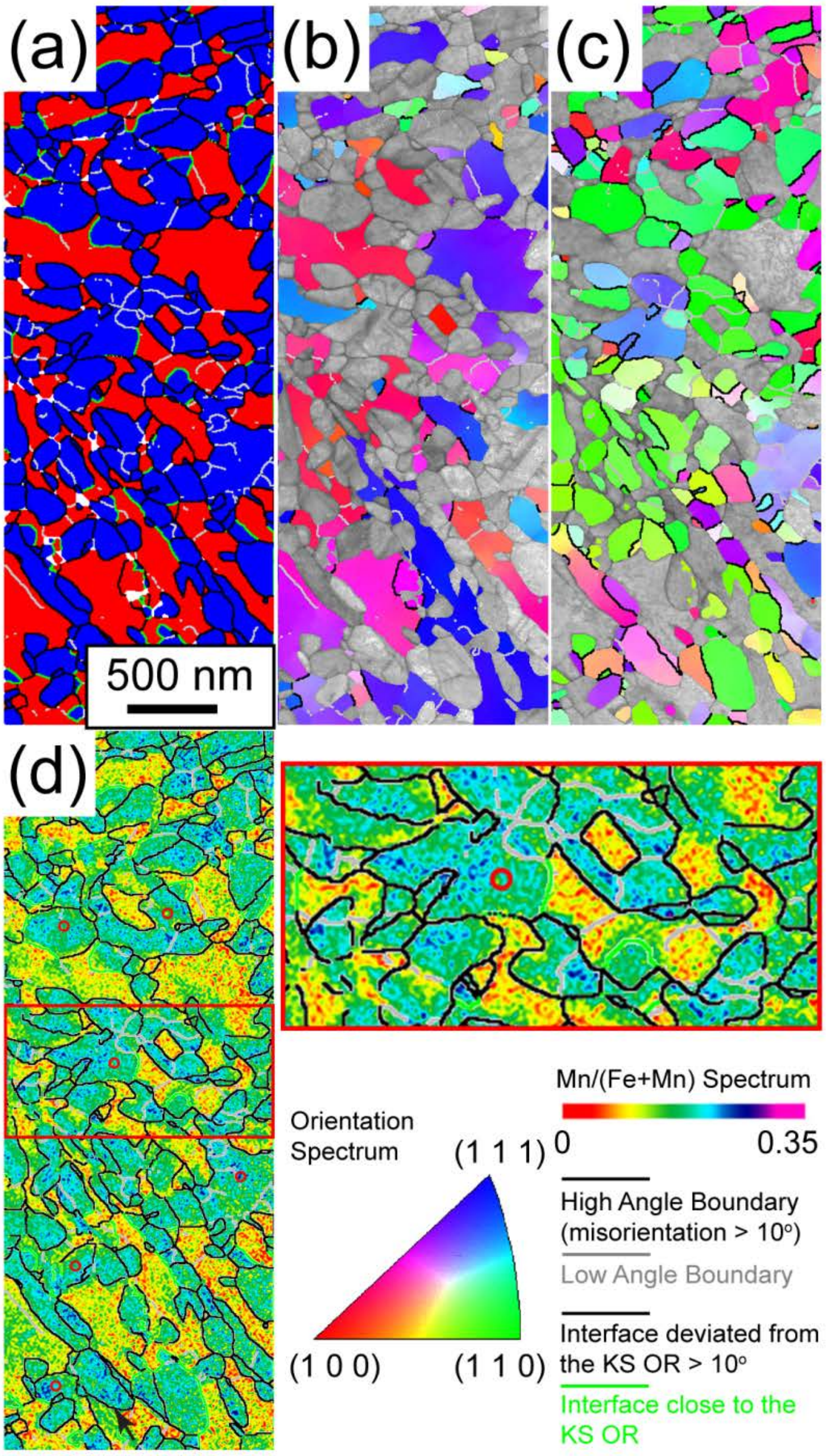

Figure 3 The SEM-TKD analyses on the 610SA steel: (a) the phase map showing ferrite (red) and austenite (blue), (b) the inverse pole figure-Z of ferrite, (c) the inverse pole figure-Z of austenite, (d) the EDXS Mn Kα /(Mn Kα +Fe Kα) peak ratio mapping.

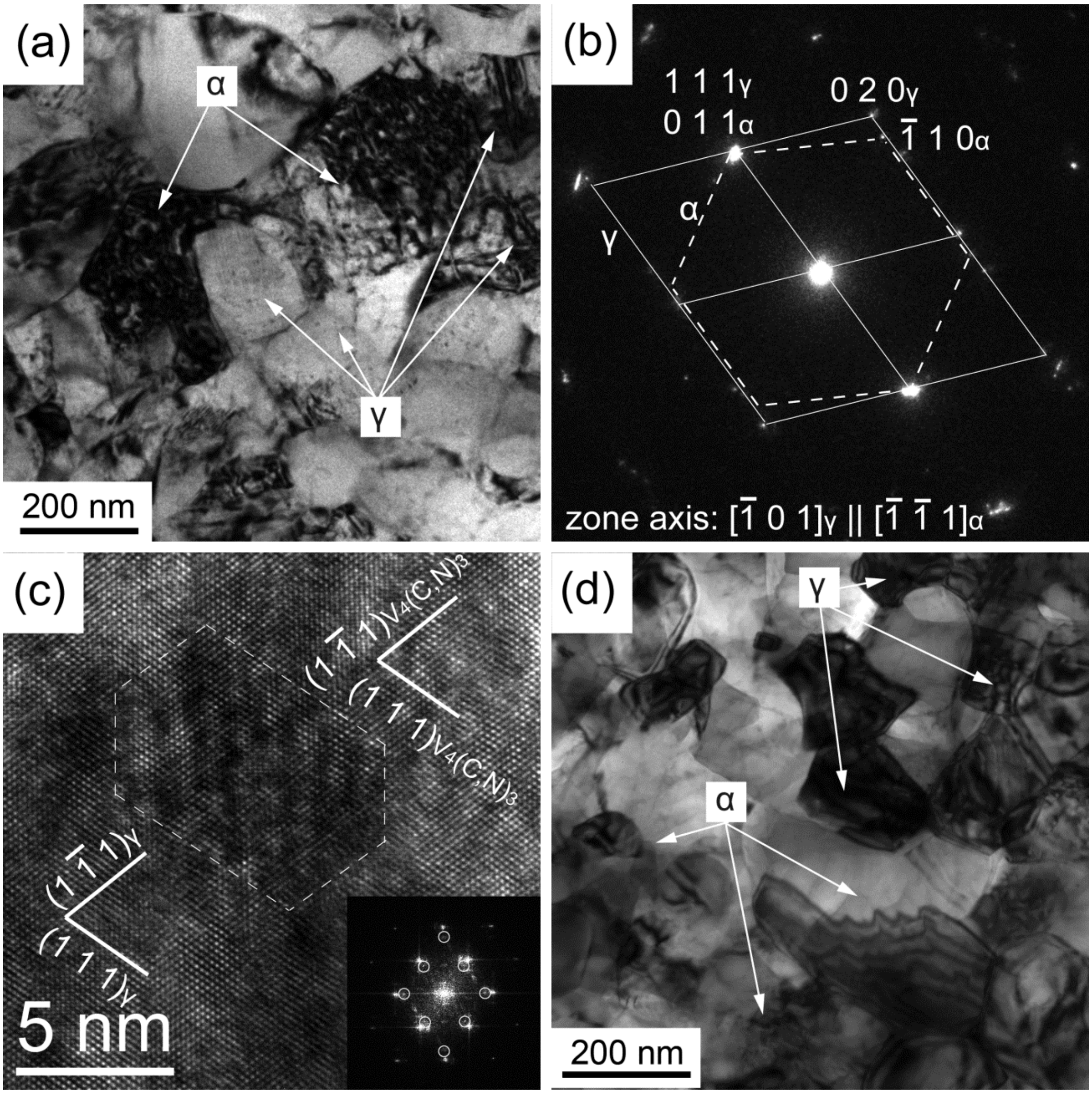

Figure 4 TEM microstructural analysis of the 550LA steel (a-c) and the 610SA steel (d). (a) Bright field images that provide an overview of the duplex microstructure, (b) the corresponding electron diffraction patterns, (c) high resolution image of $V_4(C,N)_3$ carbides in the 550LA steel, and (d) Corresponding bright field images that provide an overview of the duplex microstructure in the 610SA steel.

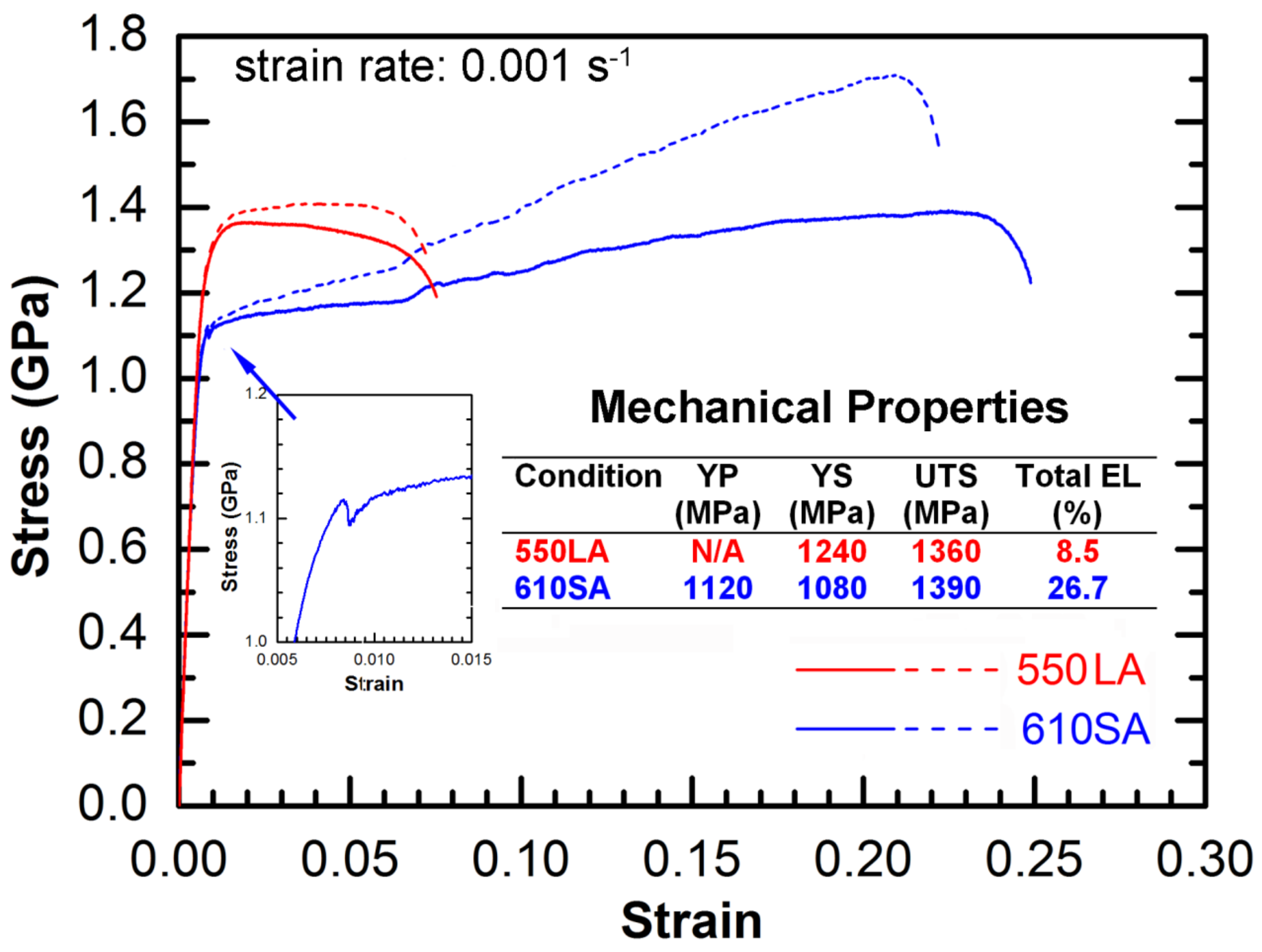

Figure 5 The engineering stress-strain curves (solid lines) and true stress-strain curves (dash lines) of the 550LA and 610SA steels.

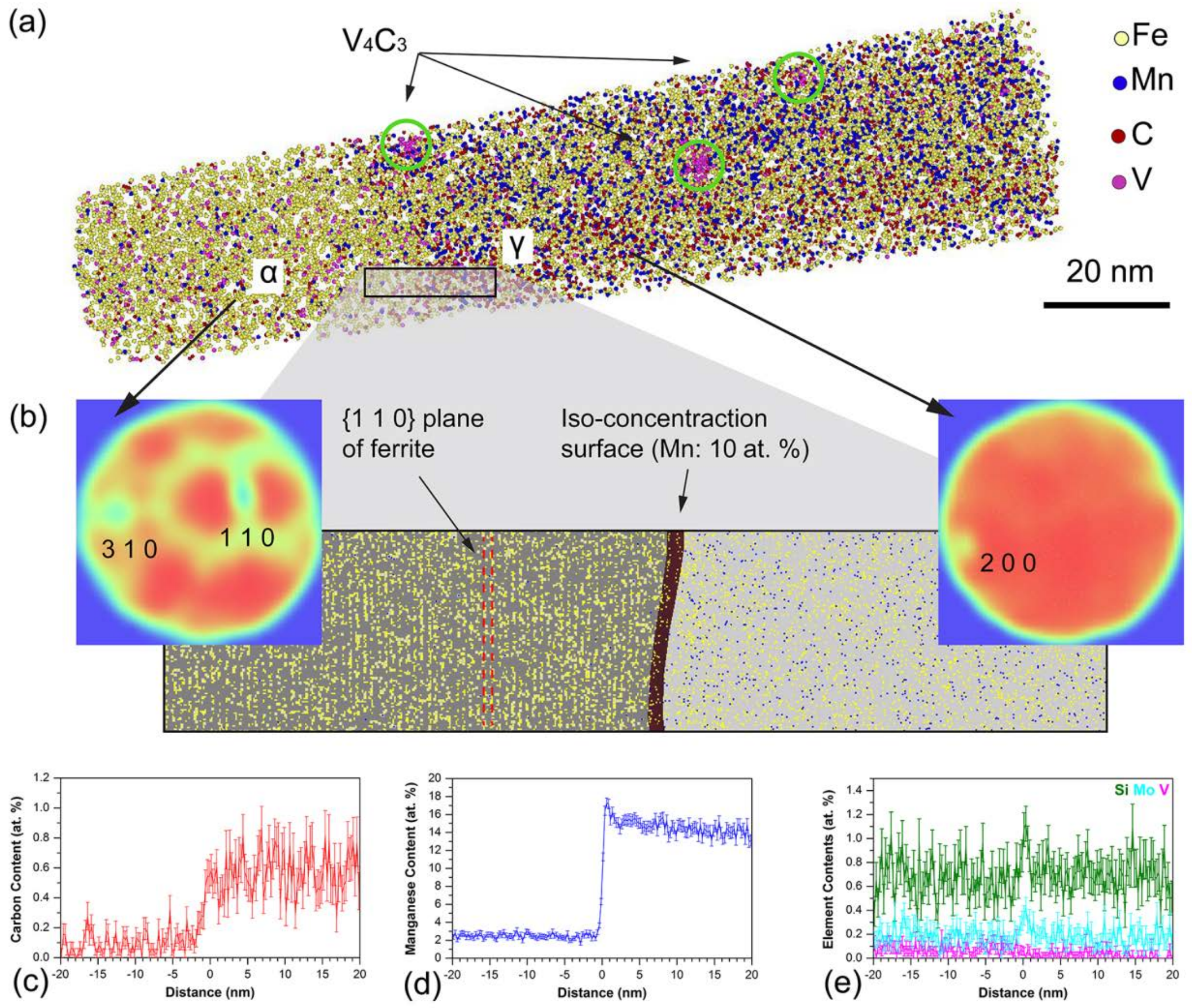

Figure 6 Atom probe tomogram of the 550LA steel: (a) atom map of iron, manganese, vanadium, and carbon, (b) iso-concentration surface of 10 at. % manganese, and compositional profiles of (c) carbon, (d) manganese, (e) silicon, molybdenum, and vanadium across the austenite/ferrite interface.

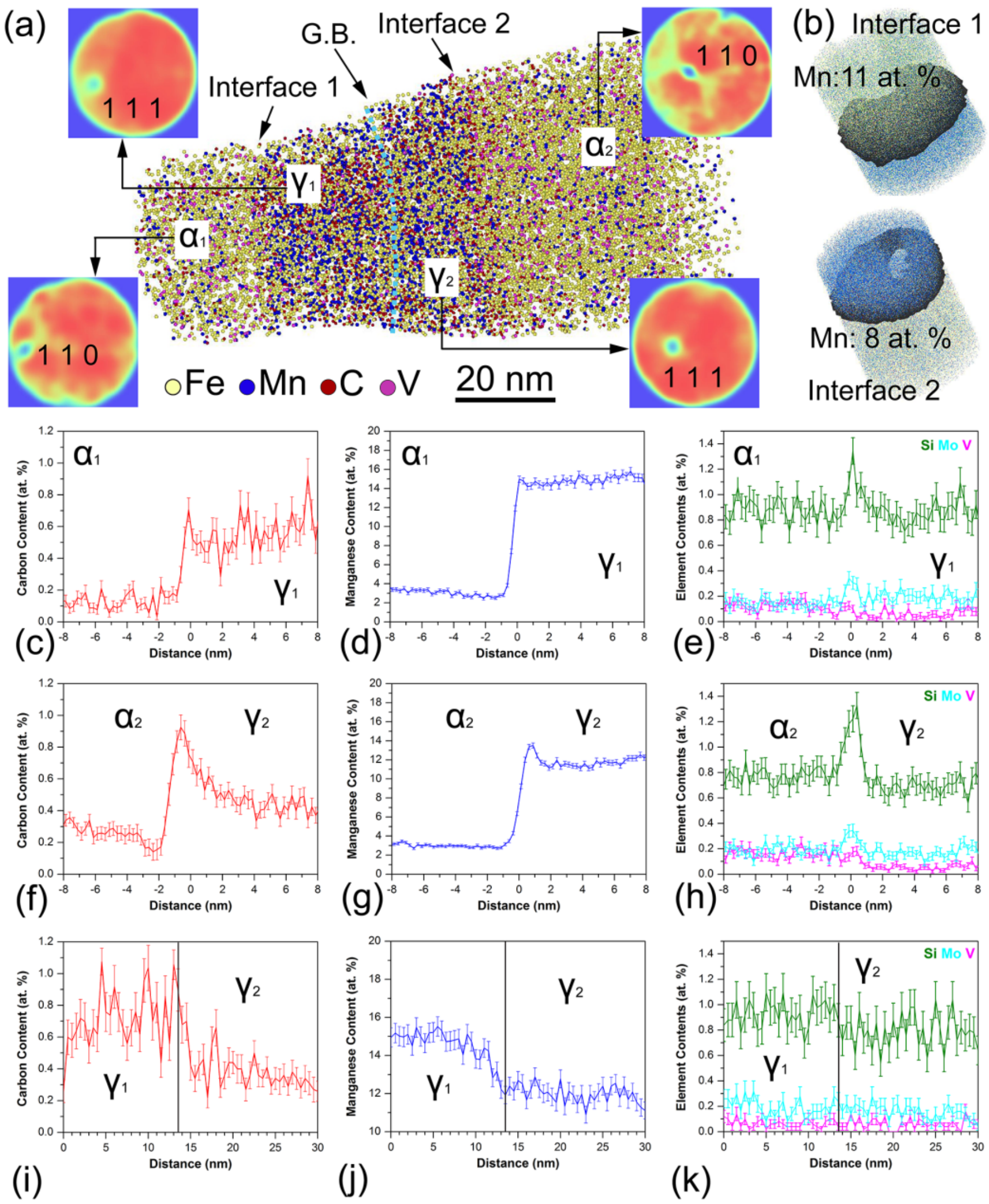

Figure 7 Atom probe tomogram of the 610SA steel: (a) atom map of iron, manganese, vanadium, and carbon, (b) iso-concentration surface of 11.0 and 8.5 at. % manganese for Interface 1 and Interface 2 respectively, and the compositional profiles of (c) carbon, (d) manganese, (e) silicon, molybdenum, and vanadium across Interface 1. The corresponding compositional profiles across Interface 2 are also provided for (f) carbon, (g) manganese, (h) silicon, molybdenum, and vanadium. The measured compositional profiles of (i) carbon, (j) manganese, (k) silicon, molybdenum, and vanadium across the austenite grain boundary are provided.

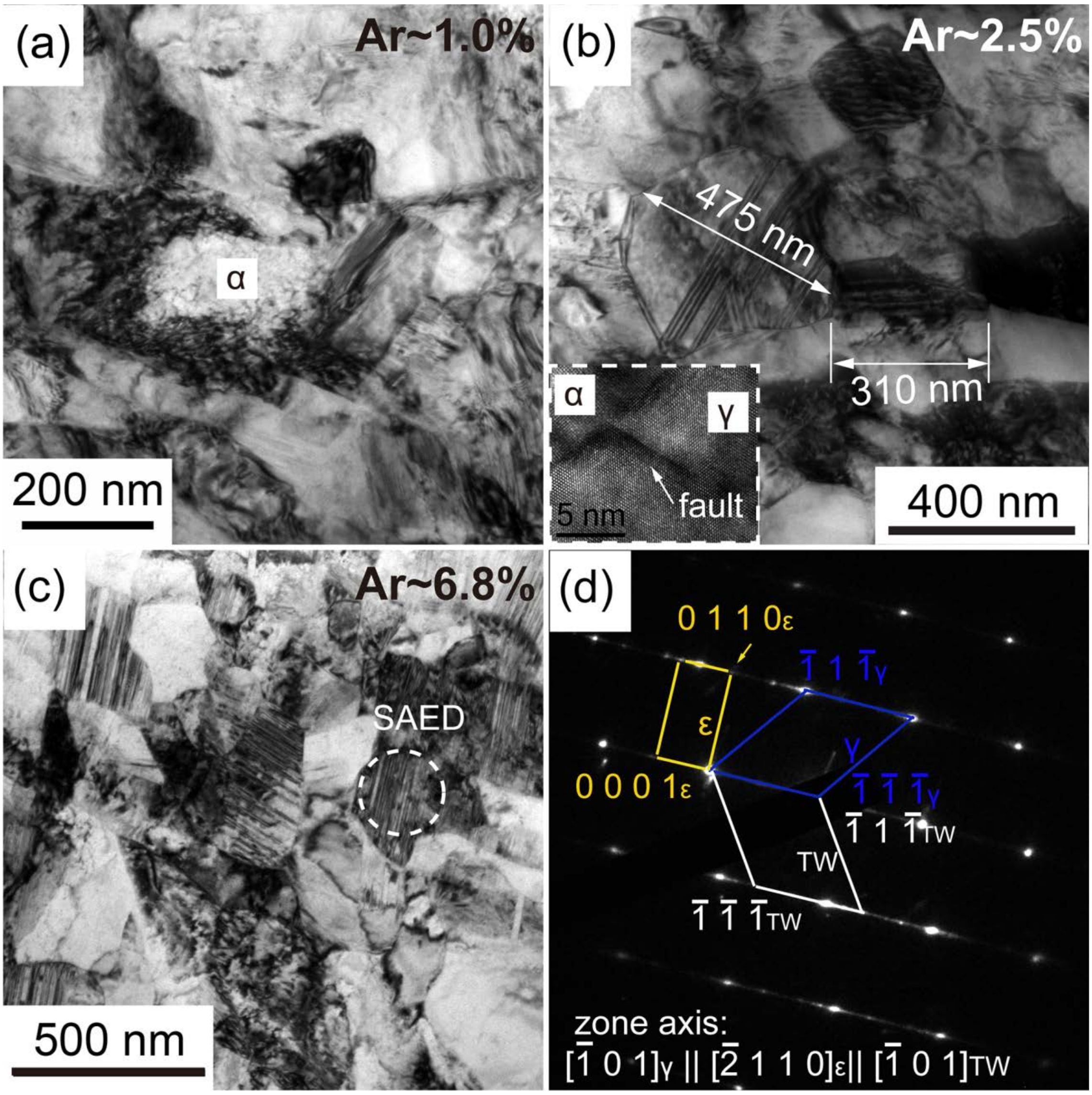

Figure 8 TEM bright field images revealing the deformation-induced microstructure in the 550LA steel strained by (a) ~1.0 %, (b) ~2.5%, and (c) ~6.8%, and (d) the corresponding electron diffraction analysis from the selected area in (c). These strain percentages relate to the percent reduction in cross-sectional area, Ar, during the uniaxial tensile test.

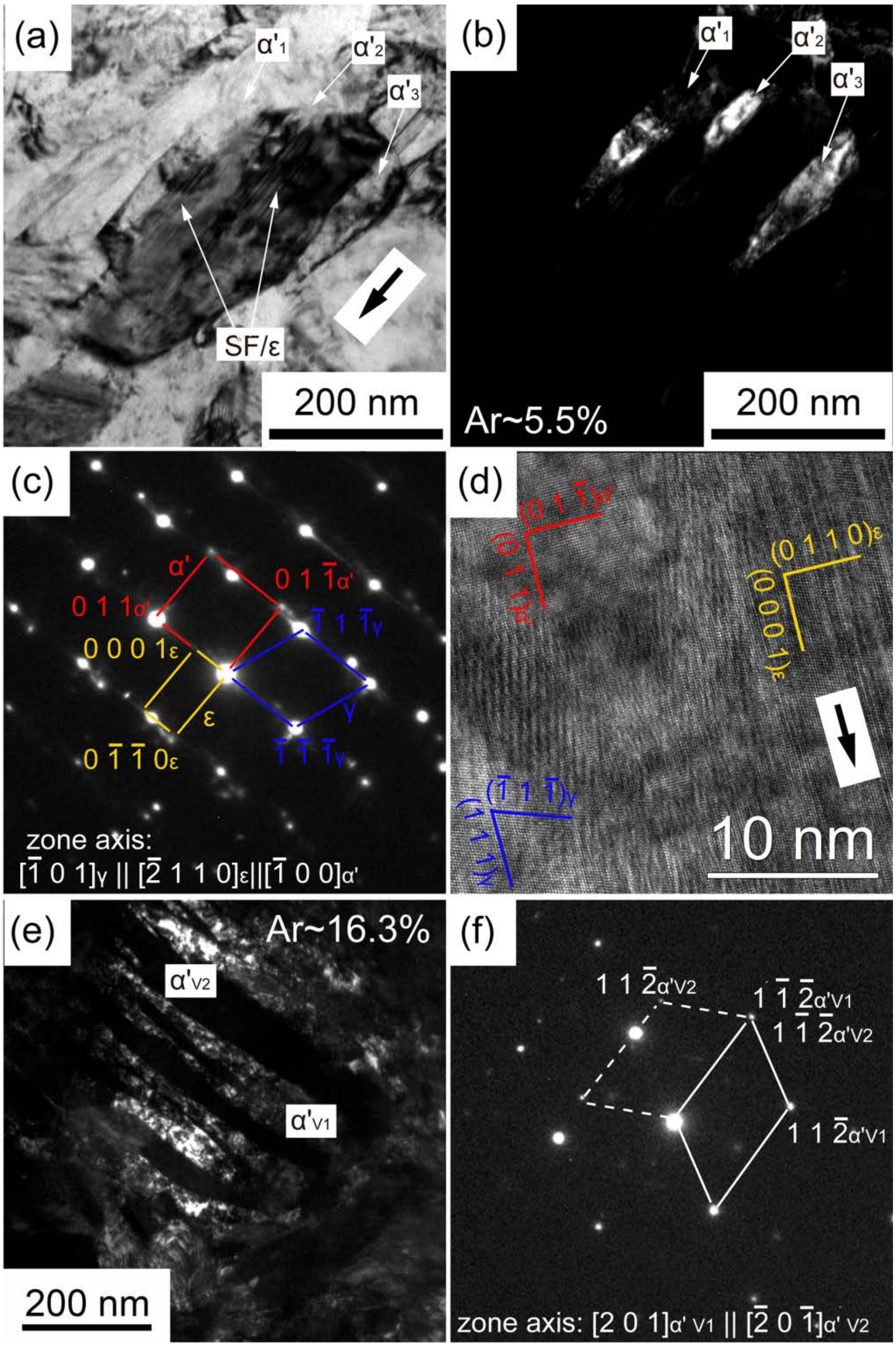

Figure 9 TEM images revealing the deformation-induced microstructure in the 610SA steel strained by 5.5 % (a)-(d) and 16.3 % (e)-(f): (a) the bright field, (b) the dark field images of α' martensite, (c) the corresponding electron diffraction analysis of (b). (d) high resolution image of the interface between α' martensite and ε martensite. (e) TEM dark field image of deformation-induced α' martensite, and (e) the corresponding electron diffraction analysis of (f).

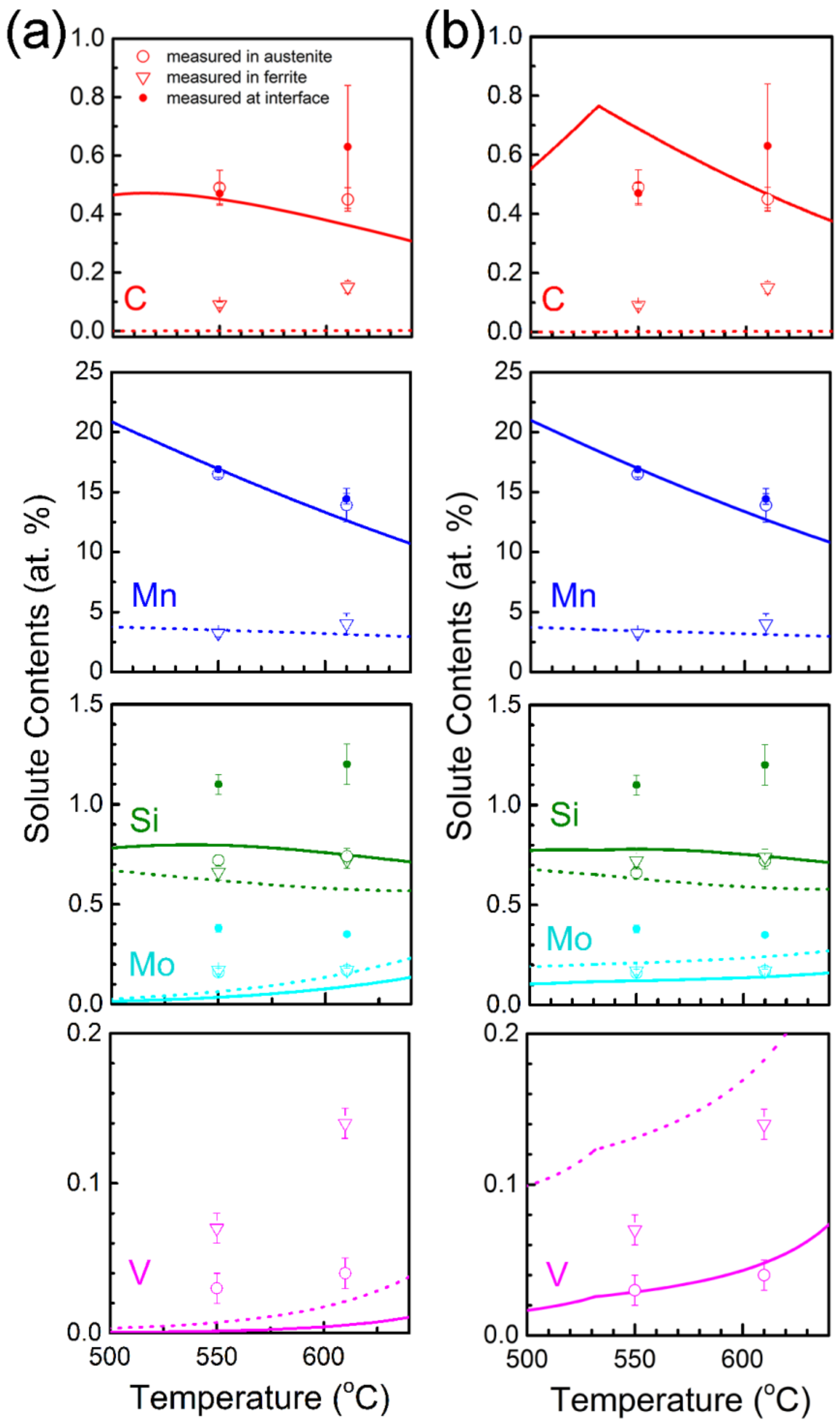

Figure 10 Comparison of atom probe microscopy results with the solute contents calculated in austenite (solid lines) and ferrite (dash lines): (a) under equilibrium conditions, and (b) under constrained conditions by suspending $Mo_2C$ and $V_4C_3$ precipitation. The data bar indicates the range of detected values.

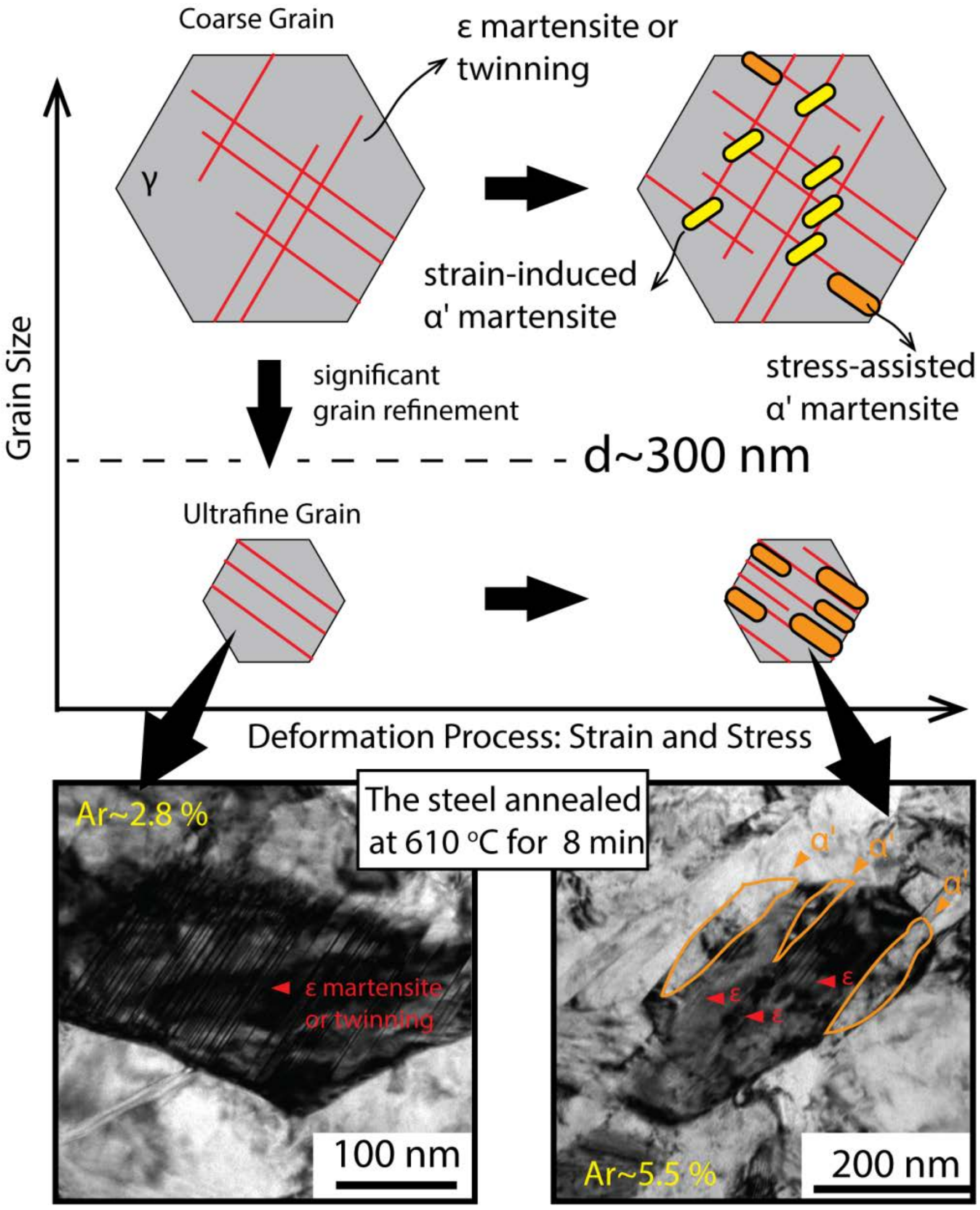

Figure 11 The schematic diagram and TEM evidence in the 610SA steel showing the relationship between grain size and mechanism of deformation-induced martensite.

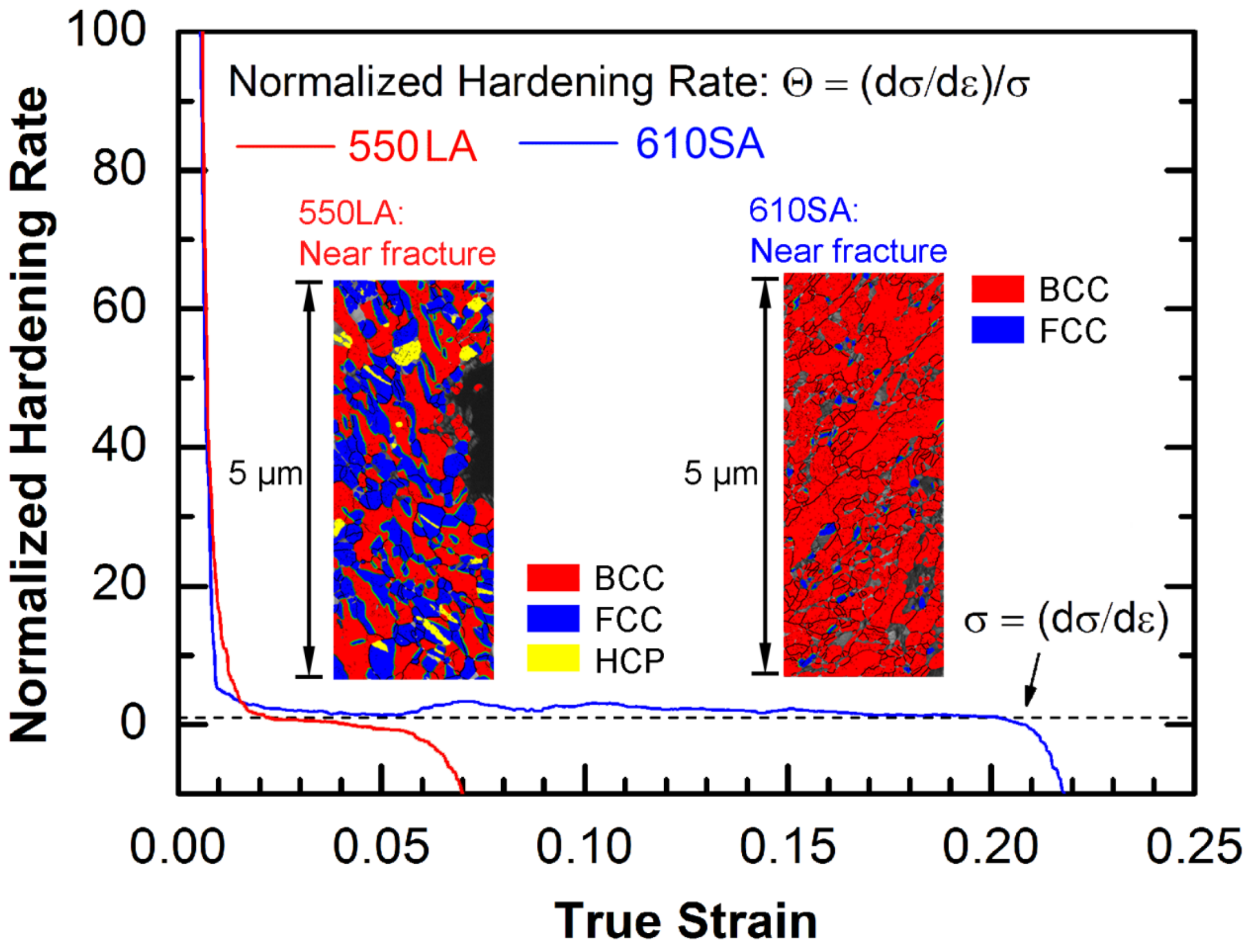

Figure 12 The change in normalized strain hardening rate with true strain and the TKD phase maps near the fracture surface of the 550LA and the 610SA steels. The black dashed line indicates the limit of plastic instability.